\documentclass[preprint,review,12pt]{elsarticle}

\usepackage{amssymb}
\usepackage{amsmath}
\usepackage{graphicx}
\usepackage{subfig}
\usepackage{multirow}
\usepackage{xcolor} 
\usepackage{bbding}

\begin{document}

\begin{frontmatter}



\title{S2LIC: Learned Image Compression with the SwinV2 Block, Adaptive Channel-wise and Global-inter Attention Context}


\author[1]{Yongqiang Wang}
\ead{wangyq0901@stu.xjtu.edu.cn}

\author[2]{Haisheng Fu}
\author[1]{Qi Cao}
\author[1]{Shang Wang}
\author[3,4]{Zhenjiao Chen}

\author[1]{Feng Liang \corref{cor1}}
\ead{fengliang@xjtu.edu.cn}
  
\address[1]{School of Microelectronics, Xi'an Jiaotong University,  Xi'an, 710049, China}
\address[2]{School of Engineering Science, Simon Fraser University, Burnaby, V5A 1S6, Canada}
\address[3]{School of Electronic Information, Northwestern Polytechnical University, Xi'an, 710072, China}
\address[4]{NO.58 Research Institute of China Electronics Technology Group Corporation, Wuxi, 214035, China}

\cortext[cor1]{Corresponding author} 


\begin{abstract}

Recently, deep learning technology has been successfully applied in the field of image compression, leading to superior rate-distortion performance. It is crucial to design an effective and efficient entropy model to estimate the probability distribution of the latent representation. However, the majority of entropy models primarily focus on one-dimensional correlation processing between channel and spatial information. In this paper, we propose an Adaptive Channel-wise and Global-inter attention Context (ACGC) entropy model, which can efficiently achieve dual feature aggregation in both inter-slice and intra-slice contexts. Specifically, we divide the latent representation into different slices and then apply the ACGC model in a parallel checkerboard context to achieve faster decoding speed and higher rate-distortion performance. We utilize deformable attention in adaptive global-inter slices context to dynamically refine the attention weights based on the actual spatial correlation and context. Furthermore, in the main transformation structure, we introduce the Residual SwinV2 Transformer model to capture global feature information and utilize a dense block network as the feature enhancement module to improve the nonlinear representation of the image within the transformation structure. Experimental results demonstrate that our method achieves faster encoding and decoding speeds, with only 0.31 and 0.38 seconds, respectively. Additionally, our approach outperforms VTM-17.1 and some recent learned image compression methods in terms of PSNR metrics, reducing BD-Rate by 8.87\%, 10.15\% and 7.48\% on three different datasets (i.e., Kodak, Tecnick and CLIC Pro). Our code will be available at https://github.com/wyq2021/S2LIC.git.

\end{abstract}




\end{frontmatter}


\section{Introduction}
\label{Introduction}

Recently, the application of deep learning to image compression has gradually outperformed traditional approaches. The primary goal of image compression is to reduce space redundancy for transmission and storage. Some traditional compression standards like JPEG \cite{jpeg}, Better Portable Graphics (BPG) \cite{bpg} and Versatile Video Coding (VVC) \cite{VVC} can effectively improve compression performance via linear transform. However, the handcrafted transformations will cause blocking effects and blurry ringing artifacts. Similar to traditional codecs, the learning-based image compression framework also includes transformations, quantization, and entropy coding. Each module consists of a trainable network in learning-based image compression architectures.

In recent years, the learned image compression (LIC) methods have developed rapidly. Some recent LIC methods \cite{fu_GLLMM,tcm_2023,tic_2022,mlic_2023,elic_2022} have outperformed the traditional VVC in terms of peak signal-to-noise ratio (PSNR) and multi-scale structural similarity (MS-SSIM). The majority of these methods are based on variational autoencoders (VAE) \cite{Ballé_2018}, which is comprised of the core autoencoder and the hyperprior coding.

In order to accurately estimate the probability distribution of the latent representation, it is crucial to design an efficient entropy model. Previous works have made significant efforts to tackle this challenge. For example, in \cite{Ballé_2018}, a scale hyperprior based on a single gaussian model is proposed, where the scale parameters are estimated using a hyperprior. Based on \cite{Ballé_2018}, Cheng \textit{et al}.\cite{Cheng_2020} have made further strides in improving the scale hyperprior by incorporating attention modules and discretized gaussian mixture module (GMM) to better parameterize latent representations, leading to significant improvements in rate-distortion performance. However, the previous methods only utilize a single distribution, resulting in spatial redundancy in the latent representation. To solve this problem, the gaussian-laplacian-logistic mixture model (GLLMM) is proposed in \cite{fu_GLLMM}. Additionally, other works have explored aspects within the context model \cite{Minnen_channelwise, elic_2022}, including the channel-wise context model and spatial context model. These  context methods lacked effective aggregation of channel-wise and spatial features, thus failing to fully utilize the correlations among these features to enhance compression efficiency. Simultaneously, there still existed redundancy within latent representations, resulting in reduced compression efficiency.


\begin{figure}[t]
\centering
\includegraphics[scale=0.55]{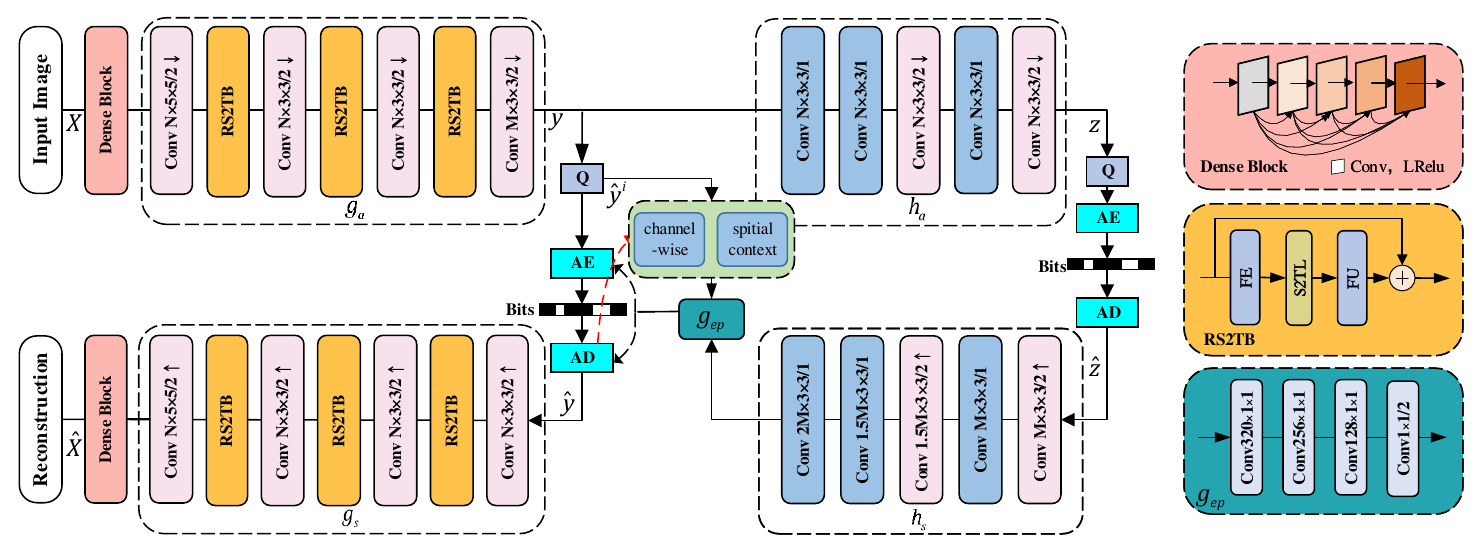}
\caption{The overall architecture of the proposed method. $g(\cdot)$ represents the analysis/synthesis transform, while $h(\cdot)$ represents the hyperprior analysis/synthesis transform. $5\times5$ and $3\times3$ indicate the sizes of the convolution kernels. $2 \uparrow$ and $2 \downarrow$ denote the up-sampling and down-sampling operations with a stride of 2. $N$ and $M$ denote the numbers of channels. $Q$ denotes quantization, while $AE$ and $AD$ stand for arithmetic encoder and arithmetic decoder, respectively. $Conv, LRelu$ refer to the convolution operation and LeakyReLU activation function.}
\label{Fig:Frame}
\end{figure}

To alleviate these limitations, we propose the adaptive channel-wise and global-inter context entropy model, which can effectively implement channel-wise and spatial feature aggregation in both inter-slice and intra-slice contexts. In our approach, the latent representation is initially divided into several slices. Each slice is further subdivided into two parts: anchor and non-anchor, which are utilized in a checkerboard context model \cite{ckbd_he} for parallel decoding. Following this, we employ an adaptive channel-wise module to extract channel context information within different slices, while applying an adaptive global-inter module across slices to model global spatial context. Furthermore, we observe that using the residual SwinV2 transformer block can significantly capture global feature information while reducing model parameters. Therefore, we aim to propose a efficient and effective model with low-latency, low-complexity and high-performance by balancing the computation complexity and compression performance. In summary, the contributions of this paper can be summarized as follows:

\begin{itemize}

\item{We propose an Adaptive Channel-wise and Global-inter attention Context model (ACGC), effectively consolidating channel and global spatial information across various slices. Moreover, we utilize deformable attention within the adaptive global-inter attention mechanism to dynamically refine attention weights, responding to spatial relationships and contexts.}

\item{We integrate ACGC into a parallel checkerboard entropy model, incorporating hyperprior side information, channel context and inter-slice global spatial information. It achieves faster decoding speed and higher rate-distortion performance. }

\item{Based upon ACGC, we further propose the S2LIC model. We adopt the Residual SwinV2 Transformer Block (RS2TB) to implement the nonlinear transformation, instead of utilizing stacked convolutional residual blocks. A feature enhancement module based on dense block concatenation is introduced before RS2TB for feature reuse and nonlinear image representation.}

\end{itemize}

Thanks for these contributions, extensive experimental results on three datasets (i.e., Kodak, Tecnick and CLIC Pro) show that the proposed method outperforms some recent works in both PSNR and MS-SSIM. Compared with VTM-17.1, the BD-rate \cite{bd_rate} was reduced by 8.87\% , 10.15\% and 7.48\% on the three datasets, respectively.

\section{Related works}
\label{Related_works}
\subsection{Learned Lossy Image Compression}
The aim of the lossy image compression is to  optimize the trade-off between rate and distortion. Giving the input image $x$ is encoded into latent representation $y$, and then $y$ is quantized into $\hat{y}$, which is decoded back to the reconstructed image $\hat{x}$ in the decoder. The basic learned image compression framework is formulated as:
\begin{equation}
\label{eq1_basic}
\hat{y} = \lceil g_{a}(x) \rfloor, \hat{x} = g_{s}(\hat{y})
\end{equation}

Where $g_{a}$ represents the analysis transform, $g_{s}$ represents the synthesis transform, and $\lceil \cdot \rfloor$ denotes the quantization operator.

In order to obtain different bit rates, we trained several independent models with different Lagrange multiplier $\lambda$ values. The optimization objective is to minimize the rate-distortion loss through end-to-end learning methods.
\begin{equation}
\label{eq2_lossfun}
\mathcal{L} = \mathcal{R}(\hat{y}) + \lambda \mathcal{D}(x, \hat{x})
\end{equation}

where $\mathcal{R}$ is the compressed bit rate of $\hat{y}$ and $\mathcal{D}$ is the distortion between the origin image $x$ and the reconstruction $\hat{x}$. The distribution of the rate $\mathcal{R}$ is the entropy $\hat{y}$, which is estimated by an entropy model during training. 

Later in \cite{Ballé_2018}, they proposed the hyperprior network to extract the side information from $y$. Adopt the hyperprior $\hat{z}$ to calculate the entropy parameter $\Theta(\mu, \sigma^{2})$. The gaussian conditional entropy model is used to estimate the rate $\hat{y}$, which can be formulated as:
\begin{gather}
\mathcal R(\hat{y}) = \mathbb{E}[-\log_2{p_{\hat{y}|\hat{z}}}({\hat{y}|\hat{z}})] \\
\mathcal{R}(\hat{z}) = \mathbb{E}[-\log_2{p_{\hat{z}}}(\hat{z})] \\
p_{\hat{y}|\hat{z}} = [\mathcal{N}(\mu,\sigma^{2}) \ast U(-\frac{1}{2},\frac{1}{2})](\hat{y})
\end{gather}

\subsection{Context-based Entropy Model}
It is crucial to design an accurate entropy model for the performance of image compression. Some current state-of-the-art entropy models mainly are comprised of channel-wise, local and global spatial attention.

Minnen \textit{et al}.\cite{Minnen_channelwise} proposed a channel-wise model. They divided the latent representation $y$ into different slices. When decoding $\hat{y}^i$, it can be conditioned on the previously decoded slice $\hat{y}^{i-1}$. However, it only considers the correlation between different channels and ignores the spatial correlation. There is a problem of uneven information distribution in different slices. ELIC \cite{elic_2022} combined the multi-dimension entropy model of space-channel context (SCCTX) into uneven slices, which can be fast and effective in reducing the bit-rate. 

Some spatial entropy contexts adopt autoregressive models \cite{fu_GLLMM, Cheng_2020} for sequential decoding, where the information to be decoded later depends on the previously decoded information. To achieve parallel decoding, He \textit{et al.} \cite{ckbd_he} divided the latent representation $\hat{y}$ into $\hat{y}_{anchor}$ and $\hat{y}_{non\_anchor}$, and proposed checkerboard convolution to extract contexts of $\hat{y}_{non\_anchor}$ from $\hat{y}_{anchor}$. Based on the transformer model and allowing for the joint learning of spatial and content information, the Entroformer model was proposed in \cite{Qian_2022}. 

Although these methods are able to capture features from multiple dimensions, there is still a lack of effective feature aggregation between channel-wise and global spatial information, and a certain correlation still exists between them. Therefore, we propose an adaptive channel-wise and global-inter attention context entropy model to achieve dual feature aggregation.

\subsection{Transformer-based Models}

Due to its excellent global feature extraction ability, transformers have achieved significant results in computer vision tasks \cite{vit}. In \cite{bai_2022}, the authors propose an end-to-end image compression and analysis model with transformers. Aiming to address global information redundancy in image compression, Qian \textit{et al.} \cite{Qian_2022} design an entropy model based on transformer instead of convolution block to predict the probability of the latent representation. A  transformer-based image compression (TIC) \cite{tic_2022} is developed, which reuses the VAE architecture with paired core and hyper encoders based on the Swin transformer \cite{tic_2022, 2021_swin}. In \cite{Li_ROI}, a region of interest (ROI) mask based on the Swin transformer block is integrated into the network architecture to provide spatial features, which achieves better ROI PSNR.

\begin{figure}[!t]
\centering
\includegraphics[width=3.5in, height=2in]{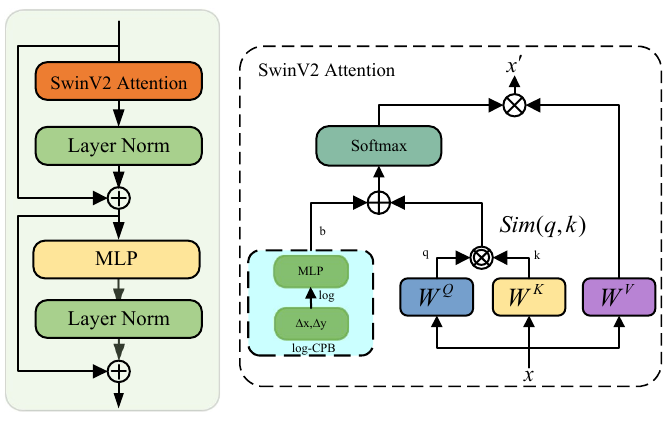}
\caption{The details of the SwinV2 Transformer Layers (S2TL) and SwinV2 Attention module. MLP refers to the multi-layer perception, while log-CPB denotes the log-space continuous position bias. Symbols $\otimes$ and $\oplus$ represent element-wise multiplication and addition, respectively.}
\label{Fig:V2_attention}
\end{figure}

In SwinV2 \cite{2022_swinV2}, the window self-attention module has been primarily modified to enhance the model's capacity and the resolution of the window. The original Swin transformer utilizes pre-normalization, which combines the output activation value of each residual module with that of the main branch. However, this will cause instability during training, as the amplitude of the main branch increases with each deeper layer. In order to effectively solve this problem, post-normalization is used in SwinV2. The output of each residual module is first normalized and then merged with the main branch. This prevents the amplitude of the main branch from accumulating layer by layer. In the original self-attention calculation, the pixel-wise attention between pairs of pixels is computed through the dot product of query and key. However, in the larger model, the attention map of certain modules and heads is primarily influenced by a limited number of pixel pairs. To alleviate this issue, the scaled cosine attention (SCA) is used. The main equation is shown as follows:
\begin{gather}
{Sim(q,k) = \frac{cosine(q,k)}{\tau}} \\
{Attention} = {Softmax}\left(Sim(q,k) + b\right) v
\end{gather}

where $q$, $k$, $v$ are the query, key and value matrices, respectively. $b$ is the relative to absolute positional embeddings obtained by projecting the position bias after re-indexing. $\tau$ is a learnable scalar that is not shared across heads and layers. And $\tau$ is set to be larger than 0.01. $Sim(q,k)$ denotes the similarity of $q$ and $k$. This block is illustrated in Fig. \ref{Fig:V2_attention}. Finally, a log-space continuous position bias method is introduced to make the relative position bias smooth across the window resolution.

\section{Methodology}
\label{Methodology}

In this section, we give a brief overview of the architecture of our model firstly, including the feature enhancement and the core transform modules. Subsequent sections will detail the checkerboard entropy module.

\subsection{Overall Architecture}
The proposed network architecture is illustrated in Fig. \ref{Fig:Frame}. The input image has a size of W $\times$ H $\times$ 3, where W, H and 3 represent the width, height, and channels of the input image, respectively. The architecture consists of three sub-networks: feature enhancement, core transformation and improved checkerboard context modules.

To further enhance the non-linear representation of the input image, we incorporate a dense block (DB) module. It is composed of five convolutional layers, each followed by a LeakyReLU activation function, with convolutional kernels measuring 3 $\times$ 3. The output of each layer is concatenated with its input to enhance the feature representation. The dense connectivity among the convolutional layers facilitates multi-level feature extraction from the input feature map, thereby enhancing the features of the input image and generating more expressive output feature maps.

The core transformation includes the analysis/synthesis transform ($g_{a}$ and $g_{s}$) and hyperprior analysis/synthesis transform ($h_{a}$ and $h_{s}$). Unlike Cheng's \cite{Cheng_2020} model, we propose a Residual SwinV2 Transformer Block (RS2TB) instead of the residual block and attention modules. The SwinV2 transformer utilizes post-normalization techniques that effectively decrease the variance of deeper features, thereby enhancing the stability of the training process. Within the RS2TB, feature embedding (FE) and feature unembedding (FU) operations adjust the input image size. Initially, the FE layer maps input features from H $\times$ W $\times$ C to HW $\times$ C dimensions. Following this, the SwinV2 Transformer Layer (S2TL) performs window-based self-attention, incorporating SwinV2 attention, layer normalization, and multi-layer perception. Ultimately, the FU layer converts the attention-enhanced features back to their original size of H $\times$ W $\times$ C.

We transform the input image $x$ into the latent representation $y$. Initially, a 5 $\times$ 5 convolutional downsampling operation is applied to minimize computational complexity and expand the receptive field. Subsequently, the data undergoes processing through a core transformation module with three layers, which includes an RS2TB and a 3 $\times$ 3 convolutional downsampling process designed to extract vital information. An entropy model network is then utilized to ascertain the probabilistic model of quantized latent representation, enabling their encoding into a bitstream. Additional details on the architecture of the entropy model will be described in the following section.

\subsection{Channel-wise Context Module}
The channel-wise context module is crucial for accurately estimating probabilities. Motivated by \cite{Minnen_channelwise} and \cite{mlic_2023}, we evenly divide the latent representation $y$ into $L$ slices $\{y^0, y^1, ..., y^L\}$, where $L$ denotes the number of slices. For the previously decoded slices $\hat{y}^{<i}$ , which can be used as the context for the current $i_{th}$ slice $y^i$, while reusing slide information to encode and decode the current slice $\hat{y}^{i}$. However, due to the quantization of the slice $y^{i}$ into $\hat{y}^{i}$, a quantization error $r=y^{i}-\hat{y}^{i}$ is inevitably generated. This quantization error leads to additional distortion in the decoded image. Therefore, we employ latent residual prediction (LRP) \cite{Minnen_channelwise} to predict this quantization error. The LRP includes a transform module with three 3 $\times$ 3 convolutional layers and utilizes the tanh activation function to scale the output appropriately, mapping it to the range (-0.5, 0.5). As the quality of the decoded slice increases, the estimation of entropy model parameters  becomes more accurate for the current slice.

\subsection{Deformable Attention for Global-inter Context Module}
The deformable attention was first proposed in \cite{cvpr_deform_2022}, they adapted deformable attention in the vision transformer and outperformed on multiple datasets. Due to its excellent performance, we apply deformable attention in learned image compression.

While channel-wise operations leverage the unique capabilities of different channels to enhance latent representation through intra-channel information exchange, capturing global spatial information within different slices is essential. Because of the global correlations between slices, we use deformable attention between the divided inter-slice. It enhances the self-attention mechanism by introducing a more flexible way of assigning attention weights. Unlike traditional self-attention module that relies on fixed positional relationships, deformable attention dynamically adjusts attention weights based on actual spatial relationships and context.  We refer to this module as the Global-inter, which extracts global information across channels from the decoded $\hat{y}^{<i}$. It enhances the self-attention mechanism by introducing a more flexible way of assigning attention weights.  

\subsection{ACGC:Adaptive Channel-wise and Global-inter Context Model}

\begin{figure}[t]
\centering
\includegraphics[scale=0.54]{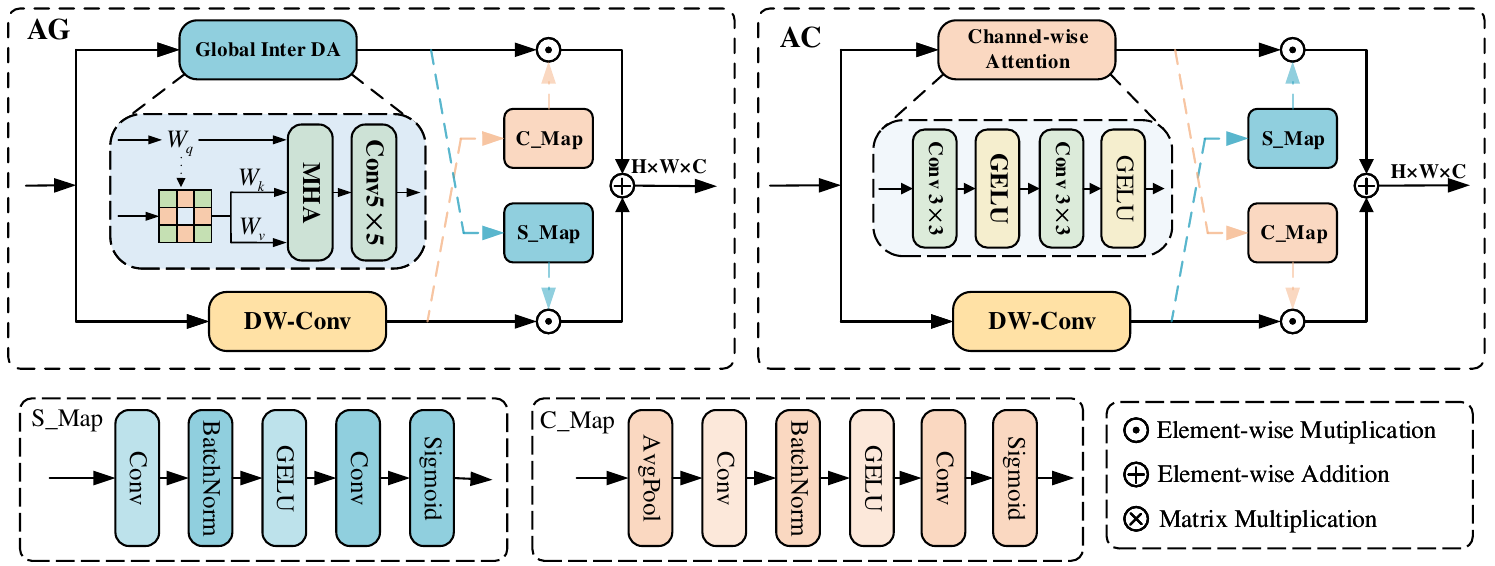}
\caption{The proposed Adaptive Channel-wise and Global-inter Context (ACGC) model. $AC$ and $AG$  refer to Adaptive Channel-wise Context and Adaptive Global-inter Context respectively. $C_{\_map}$ and $S_{\_map}$ are the channel and spatial maps in ACGC. $DW\mbox{-}Conv$ denotes Depth-wise convolution, $DA$ stands for deformable attention, $MHA$ represents multi-head attention. $Conv5\times5$ and $Conv3\times3$ indicate convolution operation with a kernel size of 5 $\times$ 5 and 3 $\times$ 3. $GELU$ refers to the GELU activation function.}
\label{Fig:ACGC}
\end{figure}

The channel-wise and global-inter context modules significantly reduce redundancy in channel and spatial information. However, focusing solely on these aspects does not fully exploit the potential correlations among slice features, which may result in some redundancy in latent representation. To further optimize the efficiency of divided slices, we aggregate features in both inter-slice and intra-slice ways between global-inter and channel-wise. Consequently, we have designed the adaptive channel-wise and global-inter (ACGC) module to reduce these redundancies. The detailed architecture of the ACGC module is shown in Fig. \ref{Fig:ACGC}.

Specifically, the ACGC module consists of two main components: the adaptive channel-wise context (AC) for channel interactions and the adaptive global-inter context (AG) for slices-inter interactions. The AG module employs deformable attention to extract feature maps from the input data and incorporates a parallel depth-wise convolution (DW-Conv). Similarly, the AC module focuses on channel-wise interactions, paralleling the approach of the AG. This dual strategy in ACGC inspired by \cite{chen_2023_dual}, optimizes the utilization of spatial and channel information, including the map operations:spatial-map ($S_{\_map}$, the size of $H\times W\times1$) and channel-map ($C_{\_map}$, with a size of $1\times1\times C$). Given the input slices features $X \in \mathbb{R}^{H\times W\times C}$, and the weight of the point-wise convolution $W_{(\cdot)}$. We can describe the operations as follows:
\begin{gather}
S_{\_map} = \sigma{(W_{2}G(W_{1}X))} \\
C_{\_map} = \sigma{(W_{2}G(W_{1}(A_{p}X)))}
\end{gather}
where $G$ denotes the GELU function, $\sigma(\cdot)$ represents the sigmiod function, and $A_{p}$ is the global average pooling. As depicted in Fig. \ref{Fig:ACGC}, the interaction process can be formulated as:
\begin{gather}
AG(G_i,D_w) = (C_{\_map}\odot G_i) \oplus (S_{\_map}\odot D_w) \\
AC(C_w,D_w) = (C_{\_map}\odot D_w) \oplus (S_{\_map}\odot C_w)
\end{gather}
where $\odot$ and $\oplus$ represent element-wise multiplication and addition, respectively. The $\odot$ represents the element-wise multiplication, $\oplus$ denotes the element-wise addition. The $G_{i}$, $C_{w}$ and $D_{w}$ correspond to global-inter, channel-wise, and depth-wise convolution operations.

\begin{figure}[!t]
\centering
\includegraphics[scale=0.58]{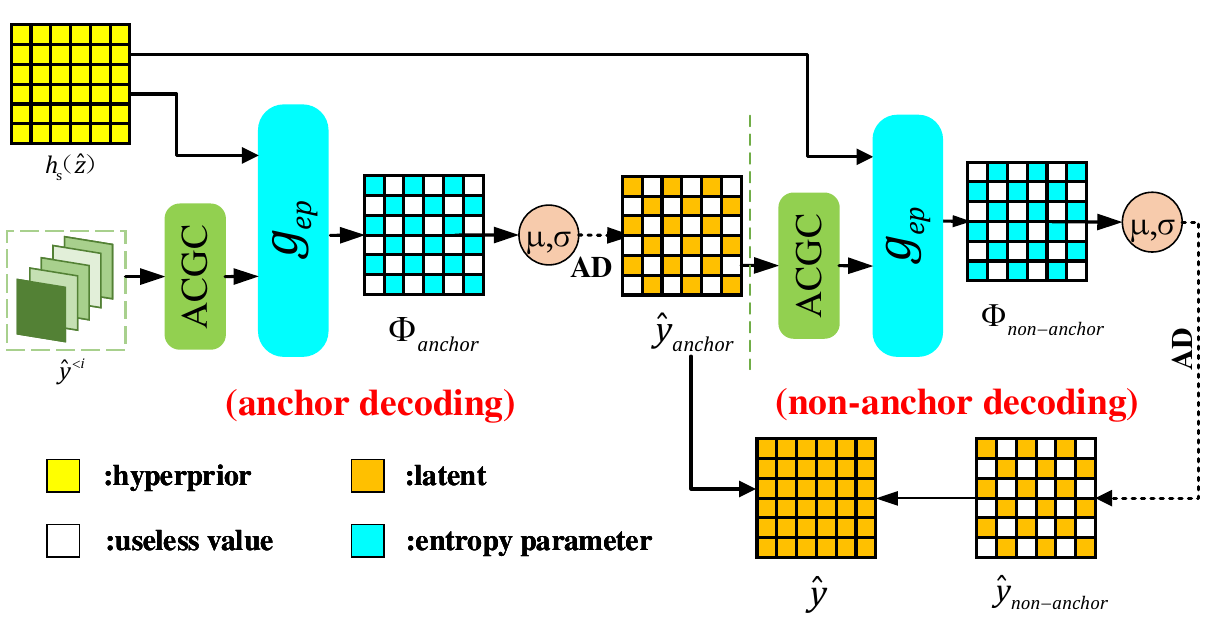}
\caption{The proposed ACGC entropy model with the checkerboard. The encoded slice $\hat{y}^{<i}$ can assist the encoding of current slice $\hat{y}$. $g_{ep}$ is the entropy parameter network.}
\label{Fig:checkerboard}
\end{figure}

As illustrated in Fig. \ref{Fig:checkerboard}, the ACGC context module is effectively utilized within the parallel checkerboard model. This setup involves inputting the hyper-parameters $\Phi_{hs}$, as well as channel $\Phi_{ch}^i$ and spatial $\Phi_{sp}^i$ information into the $g_{ep}$ network. This network predicts the entropy parameters $\Theta_i=(\mu_i, \sigma_i)$, essential for the encoding and decoding of $\hat{y}^i$ slices.

\section{Experiments}
\label{Experiments}
\subsection{Experiment Settings}
\textbf{Training Details:} Following the previous works, we select LIU4K\cite{Liu_LIU4K}, ImageNet\cite{Image_Net} and COCO2017\cite{COCO} datasets, specifically selecting images with resolutions over 480$\times$480 for training. The training process involves two phases: initially, we randomly crop images into $256 \times 256\times3$ patches for the first 1.6M steps, and subsequently for larger images (minimum 448 pixels in width and height), we crop them into $448 \times 448 \times 3$ patches. The proposed model is implemented on the open-source CompressAI PyTorch library \cite{compressai}. All the experiments are conducted on RTX 4090 GPU for 500 epochs. The training process utilizes an initial learning rate of $10^{-4}$, using the Adam \cite{Adam} optimizer with hyper-parameters $\beta_{1}=0.9$, $\beta_{2}=0.999$. Additionally, the batch size is set to 8.

We use the mean squared error (MSE) and MS-SSIM \cite{ms-ssim} as quality metrics to optimize our models. For the MSE, the parameter $\lambda$ is chosen from the set of \{0.0018, 0.0035, 0.0075, 0.013, 0.025, 0.048\}. While for the MS-SSIM, the $\lambda$ is the set of \{5,16,36,64,80\}. The number of channels is set to \emph{N} = 192 and \emph{M} = 320 for training. The other parameters follow the setting in \cite{tic_2022}.

\textbf{Evaluation:} The test datasets are the Kodak \cite{Kodak}, Tecnick \cite{Tecnick} and CLIC professional validation datasets \cite{clic}. The Kodak dataset consists of 24 images with a resolution of 512 $\times$ 768 or 768 $\times$ 512. The Tecnick dataset contains 100 high-resolution images, each sized at 1200 $\times$ 1200. As for the CLIC professional validation (CLIC Pro) dataset, which is comprised of 41 high-quality images with 2K resolution. We evaluate our model with some recent learned image compression methods and some traditional image codecs by using the PSNR and the MS-SSIM \cite{ms-ssim}.

\begin{table}[ht]
\centering 
\caption{The complexity comparison results for recent works on the Kodak dataset. Enc.Total and Dec.Total denote total time for encoding and decoding respectively.}
\label{Table:complexity}
\setlength{\tabcolsep}{13pt}
\begin{tabular}{ccc}
\hline
\multirow{2}{*}{Methods} & \multicolumn{2}{c}{Kodak \cite{Kodak} $(768\times512)$} \\
\cline{2-3}
& Enc.Total (s) & Dec.Total (s) \\
\hline
VTM-17.1  & 402.27	& 0.60	   \\ 
\hline
Cheng'20(CVPR'20) \cite{Cheng_2020} & 5.52	& 13.68	   \\ 
\hline
Xie'21(ACMMM'21) \cite{xie_2021_enhanced}& 3.86	&9.82  \\ 
\hline
Entroformer(ICLR'22) \cite{Qian_2022}& 18.78	& 0.95  \\ 
\hline
WACNN(CVPR'22)   \cite{wacnn_cvpr_2022}& 0.21	& 0.24	  \\ 
\hline
ELIC(CVPR'22) \cite{elic_2022}& 0.49	& 0.21		\\ 
\hline
TIC(DCC'22) \cite{tic_2022} & 8.18	& 14.59	   \\ 
\hline
GLLMM'23(TIP'23) \cite{fu_GLLMM}& 467.90	& 467.90	  \\ 
\hline
Fu'23(TCSVT'23) \cite{fu_TCSVT_2023}& 22.62 & 23.51 \\
\hline
MLIC(ACMMM'23)  \cite{mlic_2023}& 0.56	& 0.60 \\ 
\hline
\textbf{S2LIC(Ours)} & \textbf{0.31}  & \textbf{0.38}	\\ 
\hline
\end{tabular}
\end{table}

\subsection{Complexity Analysis}
In S2LIC, we use a parallel checkerboard context model for ACGC and the latent representation $y$ is divided into ten slices. We emply a NVIDIA GTX 2080Ti GPU and 2.9GHz Intel Xeon Gold 6226R CPU to evaluate. On the complexity of encoding and decoding time in Table \ref{Table:complexity}, we compare our model with other recent methods. As can be seen, VTM-17.1 takes the longest to encode, but it decodes quite quickly once encoded, requiring only 0.6 seconds. Cheng'20\cite{Cheng_2020} ,Fu'23\cite{fu_TCSVT_2023} and GLLMM\cite{fu_GLLMM} used an autoregressive context model for entropy coding, resulting in longer decoding time. Despite achieving state-of-the-art performance, the GLLMM\cite{fu_GLLMM} model's speed is notably affected by its increased complexity. Our model has showcased remarkable results, with encoding and decoding time of only 0.31 and 0.38 seconds, respectively. 

\subsection{Rate-Distortion Performance}

\begin{figure}[!t]  
\centering  
\subfloat{  
\begin{minipage}[b]{0.48\linewidth}  
\centering  
\includegraphics[width=\columnwidth]{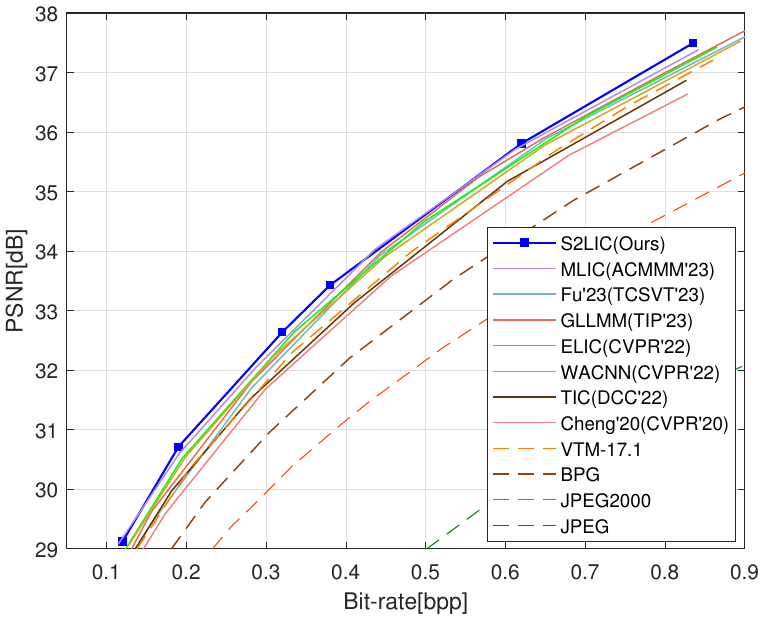}  
\end{minipage}  
}%
\subfloat{  
\begin{minipage}[b]{0.48\linewidth}  
\centering  
\includegraphics[width=\columnwidth]{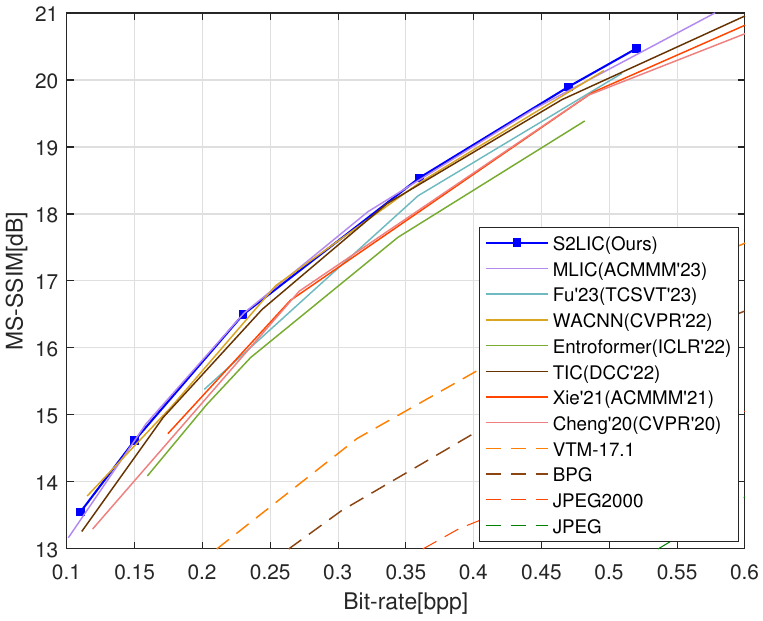}  
\end{minipage}  
}%
\caption{Rate-Distortion curves of various comparison results on all 24 Kodak images in terms of PSNR and MS-SSIM.}  
\label{Fig:keda}  
\end{figure}  
\begin{figure}[!t]  
\centering  
\subfloat{  
\begin{minipage}[b]{0.48\linewidth}  
\centering  
\includegraphics[width=\columnwidth]{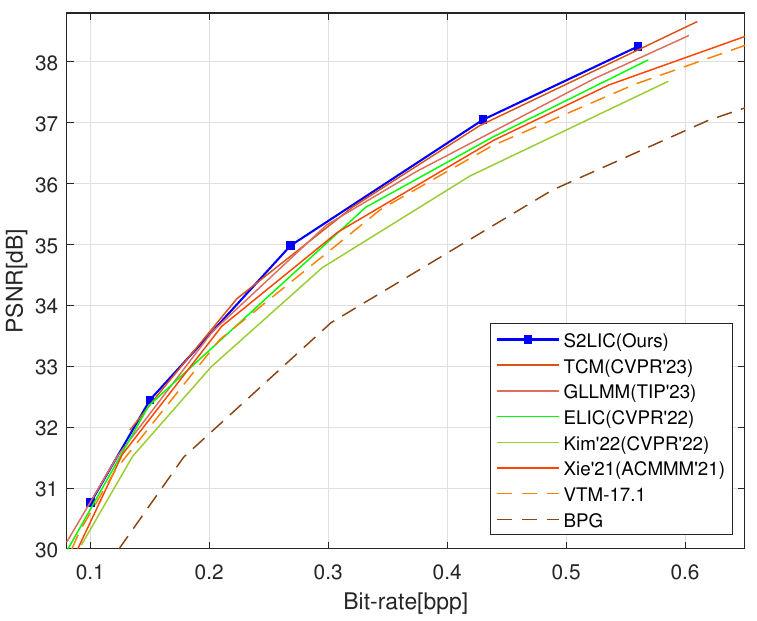}  
\end{minipage}  
}%
\subfloat{  
\begin{minipage}[b]{0.48\linewidth}  
\centering  
\includegraphics[width=\columnwidth]{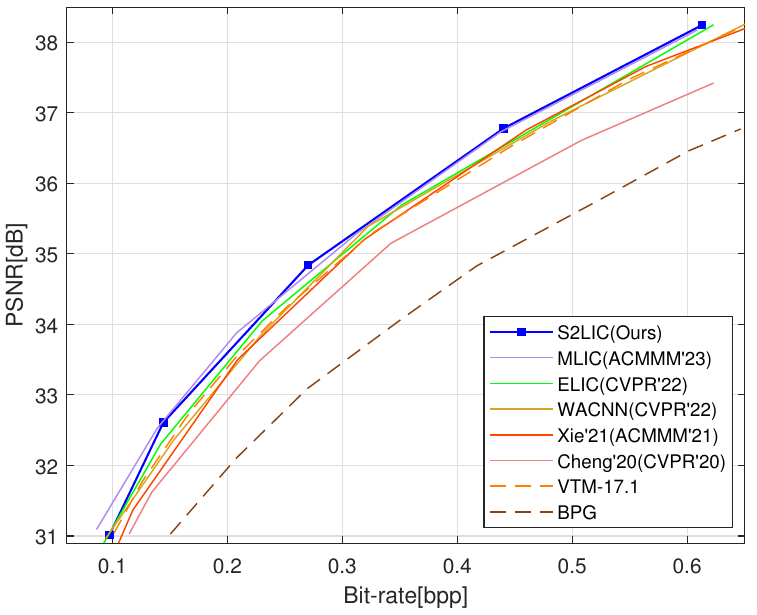}  
\end{minipage}  
}%
\caption{Rate-Distortion curves of various comparison results on Tecnick images and CLIC Pro images in terms of PSNR.}  
\label{Fig:Tecnick_CLIC}  
\end{figure}

\begin{table}[ht]
\centering
\caption{BD-Rate(\%) comparison for different models in terms of PSNR (dB) and MS-SSIM (dB) on three datasets. We use the VTM-17.1 intra as the anchor (BD-Rate=0.00\%). When the comparison model shows better results than anchor BD-Rate value less than 0\%. ``$--$'' means the result is not available due to the lack of relevant comparative results from these models.}
\label{Table:bd_rate}
\resizebox{\linewidth}{!}{
\begin{tabular}{ccccccc}
\hline
\multirow{2}{*}{Methods} & \multicolumn{2}{c}{Kodak} & \multicolumn{2}{c}{Tecnick} & \multicolumn{2}{c}{CLIC Pro} \\
{} & PSNR & MS-SSIM & PSNR & MS-SSIM & PSNR & MS-SSIM  \\
\hline
VTM-17.1  & 0.00	&  0.00	& 0.00	& 0.00	& 0.00	& 0.00  \\ 
\hline
BPG \cite{bpg} & +20.23	& +23.73	& +36.93	& +28.68	& +39.91	& +39.63  \\ 
\hline
Cheng'20(CVPR'20) \cite{Cheng_2020}  & +3.79 & -47.05	& +3.58	& -40.41 & +11.20	& -41.73   \\
\hline
Xie'21(ACMMM'21) \cite{xie_2021_enhanced} & -4.38 & -45.41	& -3.19	& $--$	& -1.63	& $--$    \\ 
\hline
TIC(DCC'22) \cite{tic_2022} & +0.32	& -49.62	& $--$	& $--$	& $--$	& $--$   \\ 
\hline
Entroformer(ICLR'22) \cite{Qian_2022} & -0.07	& -45.41	& +0.42	& $--$	& $--$	& $--$  \\ 
\hline
WACNN(CVPR'22) \cite{wacnn_cvpr_2022}  & -6.48	& -49.75	& $--$	& $--$	& -1.07	& -44.71 \\ 
\hline
ELIC(CVPR'22) \cite{elic_2022} & -5.47	& -54.54	& -6.23	& $--$	& -3.49	& $--$ \\ 
\hline
Fu'23(TCSVT'23) \cite{fu_TCSVT_2023} & -5.28	& -47.07	& $--$	& $--$	& $--$	& $--$   \\ 
\hline
TCM'23(CVPR'23) \cite{tcm_2023}& -6.78	& -49.69	& -6.07 & $--$	& $--$ & $--$\\ 
\hline
GLLMM'23(TIP'23) \cite{fu_GLLMM}& -7.39	& -49.69	& -9.53 & -46.51	& $--$	& $--$ \\ 
\hline
MLIC'23(ACMMM'23) \cite{mlic_2023}& -8.11	& -49.25	& -9.72 & $--$	& -6.93	& $--$ \\   
\hline
\textbf{\textcolor{blue}{S2LIC(Ours)}} & \textbf{\textcolor{blue}{-8.87}} & \textbf{\textcolor{blue}{-50.39}} & \textbf{\textcolor{blue}{-10.15}}	& \textbf{\textcolor{blue}{-47.28}}	& \textbf{\textcolor{blue}{-7.48}}	& \textbf{\textcolor{blue}{-45.53}}	\\ 

\hline
\end{tabular}	   	   	   
}
\end{table}

In this section, we compare our S2LIC model with recent state-of-the-art (SOTA) learned image compression models. The traditional image compression codecs, including VTM-17.1\cite{VVC}, BPG\cite{bpg}, JPEG2000 and JPEG are evaluated in terms of both PSNR and MS-SSIM metrics. For a clearer comparison, we convert MS-SSIM values to $-10\log_{10}(1-\text{MS-SSIM})$. The rate-distortion performance on the Kodak dataset is shown in Fig. \ref{Fig:keda}. Our method surpasses VTM-17.1 in PSNR and demonstrates a 0.3 to 0.6 dB improvement over GLLMM\cite{fu_GLLMM}. The performance on the Tecnick and CLIC Pro datasets is detailed in Fig. \ref{Fig:Tecnick_CLIC}, showcasing similar performance. Furthermore, we present the BD-Rate\cite{bd_rate} as the quantitative metric for the Kodak, Tecnick and CLIC Pro datasets in Table \ref{Table:bd_rate}, with VTM-17.1 as the anchor (BD-Rate = 0\%). Our S2LIC reduces the BD-Rate by 8.87\%, 10.15\% and 7.48\% on these datasets when measured in PSNR.

\begin{figure}[t]
\centering
\includegraphics[width=4.5in]{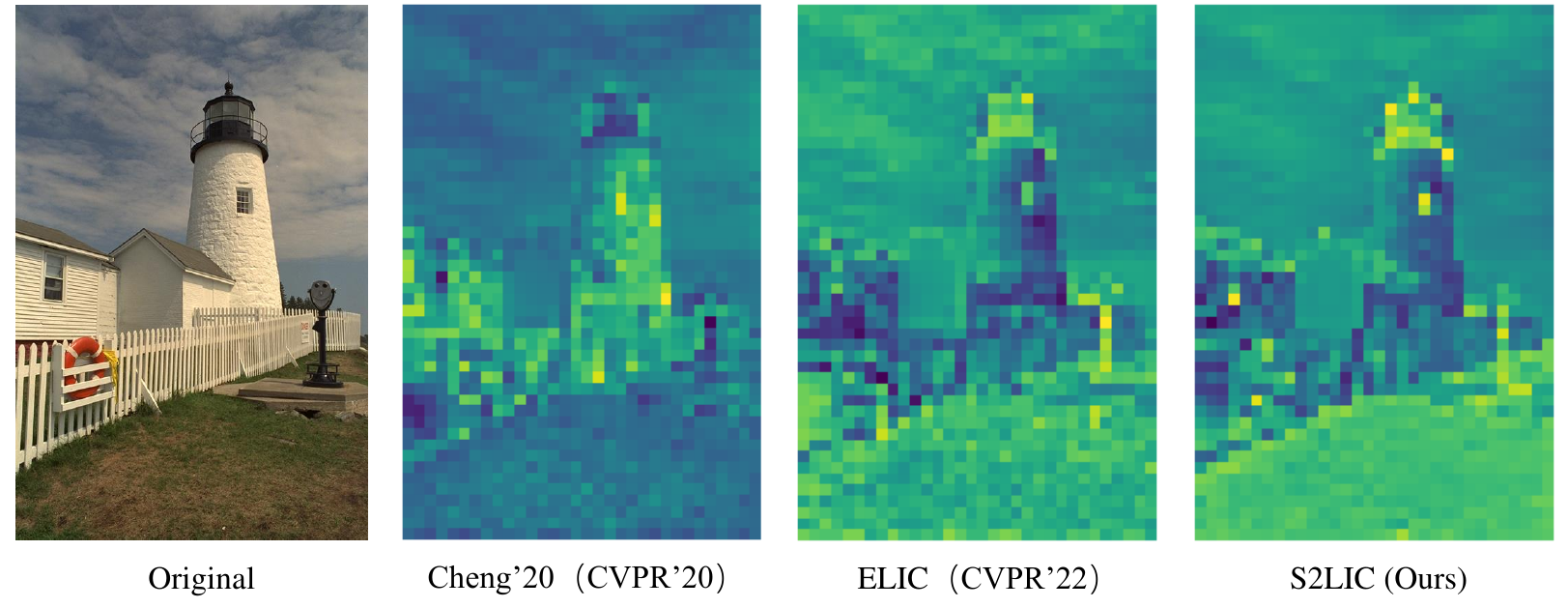}
\caption{Visualization of the average latent feature maps $\hat{y}$ of $kodim\_19$ from the Kodak dataset on different models. The compared models include Cheng'20(CVPR'20)\cite{Cheng_2020} and ELIC(CVPR'22)\cite{elic_2022} (optimized for MSE).}
\label{Fig:v_latent}
\end{figure}

\subsection{Qualitative Results}
\begin{figure*}[t]
\centering
\subfloat[Original]{\includegraphics[width = 0.33\textwidth]{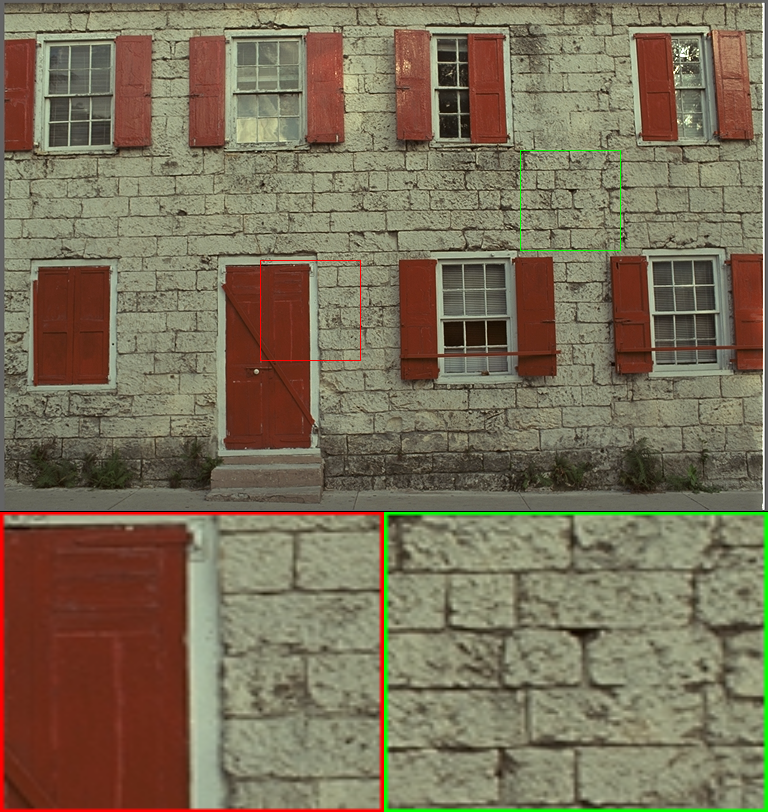}}
\hfill
\subfloat[Cheng'20(0.174/25.80)]{\includegraphics[width = 0.33\textwidth]{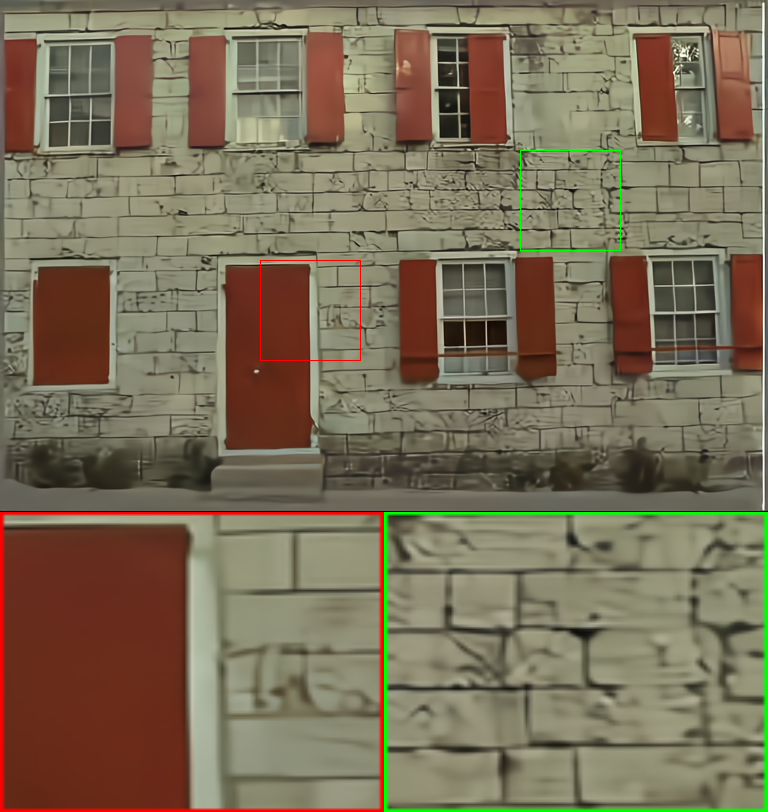}}
\hfill
\subfloat[ELIC'22(0.171/25.97)]{\includegraphics[width = 0.33\textwidth]{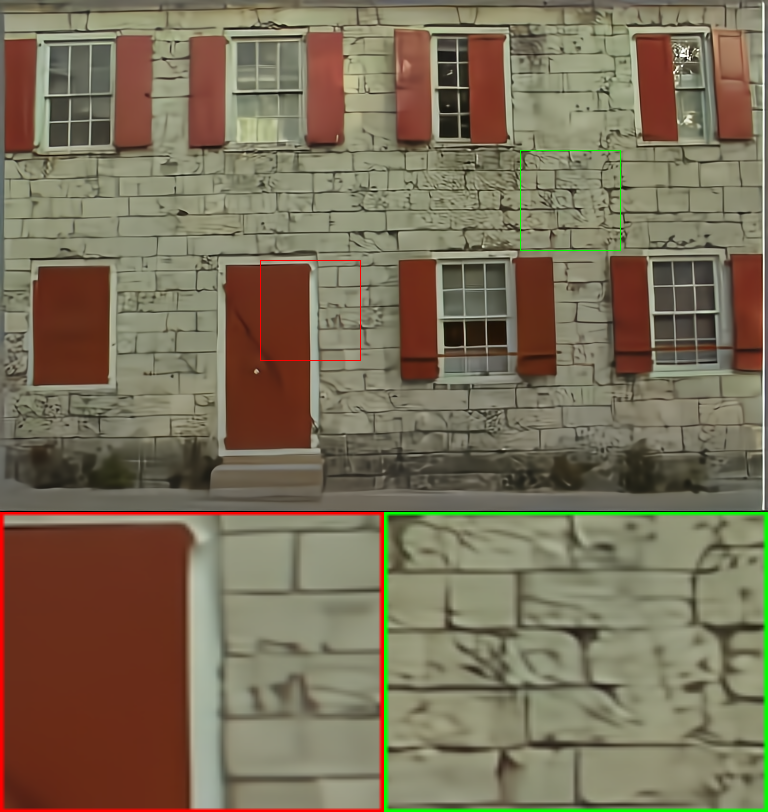}} 
\hfill
\subfloat[BPG(0.192/25.51)]{\includegraphics[width = 0.33\textwidth]{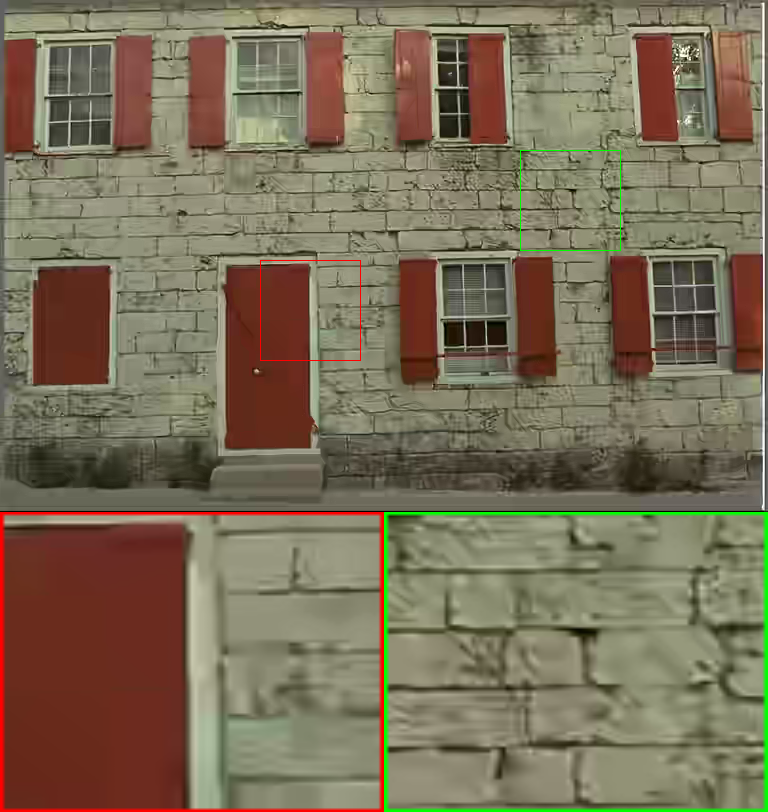}}
\hfill
\subfloat[VTM-17.1(0.167/25.83)]{\includegraphics[width = 0.33\textwidth]{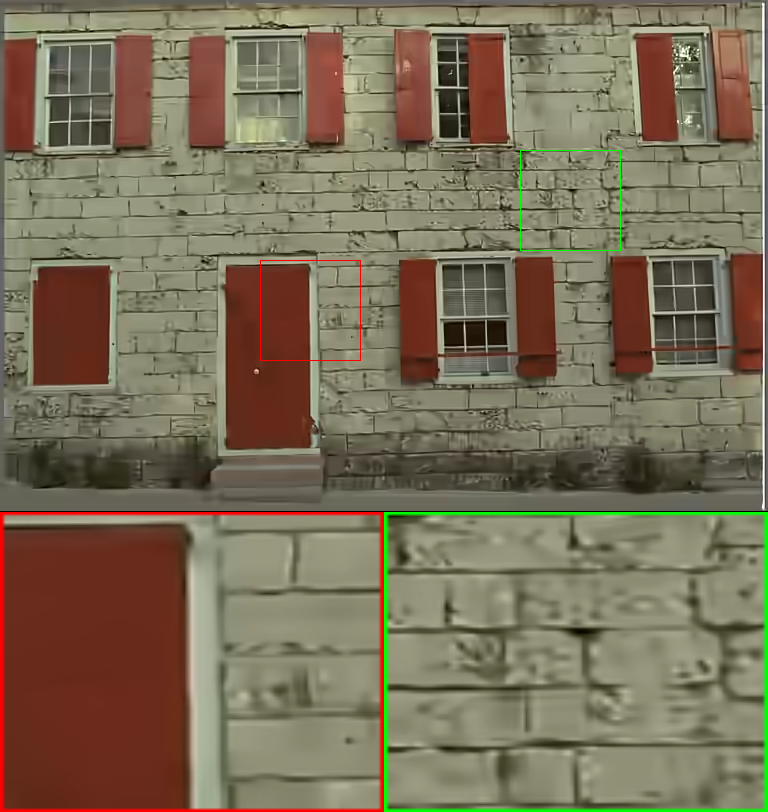}} 
\hfill
\subfloat[Ours(0.178/26.13)]{\includegraphics[width = 0.33\textwidth]{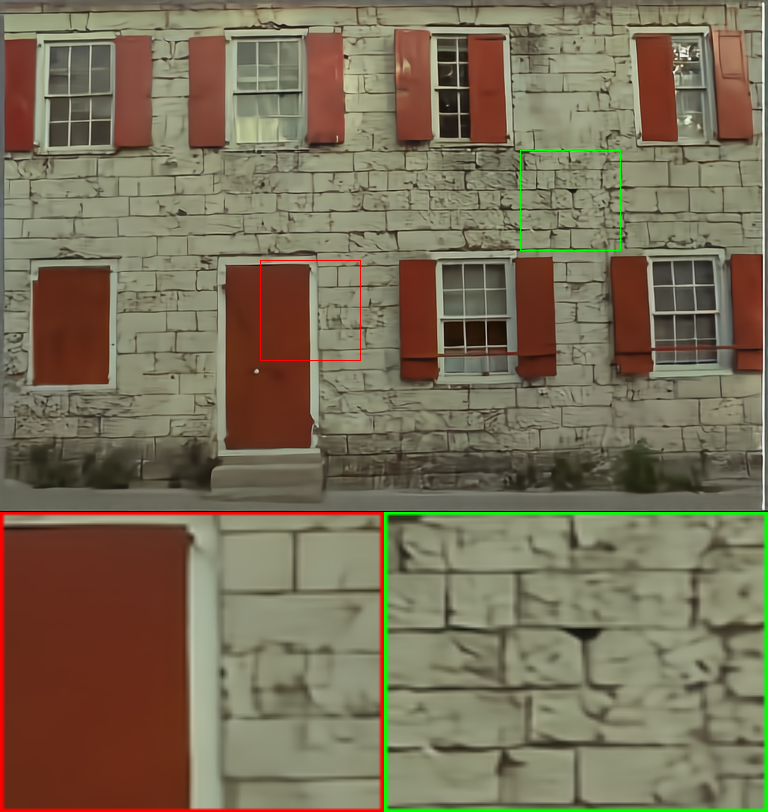}} 
\caption{Comparison of the visual reconstructed $kodim\_01$ image in Kodak dataset on different models. The metrics are [bpp/PNSR].}
\label{Fig:Q_1}
\end{figure*}

We select $kodim\_01$ images from the Kodak dataset as evaluation samples for a qualitative comparison. Fig. \ref{Fig:Q_1} illustrates visual comparisons of reconstructed images by various models, including Cheng'20\cite{Cheng_2020}, ELIC\cite{elic_2022}, VTM-17.1 and BPG. For a detailed observation and comparison, the lowest bitrate was chosen. Notably, our method generates more details in the reconstructed images, making them visually more similar to the original images.

Fig. \ref{Fig:v_latent} shows the average representation of latent feature maps.  We compared two models of Cheng'20\cite{Cheng_2020} and ELIC\cite{elic_2022}(optimized by MSE). They use the same attention in the analysis transform module, with the difference being that Cheng'20 employing GMM probability model and ELIC utilizing SCCTX model. In our model, we replace the attention module with RS2TB and utilize the proposed ACGC in the entropy module. The S2LIC feature maps effectively capture local characteristics, such as the positions of the top tower and windows. Additionally, the edges of the image are clearer. Simple regions like the sky and grass have lower energy concentration in feature maps, suggesting that fewer bits are allocated to these areas.

\subsection{Ablation Studies}
In order to compare different components and further verify the contributions of the context module and analysis transform module on performance, we conduct the corresponding ablation studies. Similar to the previous experiment, we train for 200 epochs on the LIU4K\cite{Liu_LIU4K} and COCO2017\cite{COCO} datasets. During ablation studies, we crop images into $256\times256\times3$ patches. The initial learning rate is set to $10^{-4}$ with a batch size of 8.

\begin{table}[!t]
\caption{Ablation study of ACGC module. The anchor is VTM-17.1 intra (BD-Rate =0.00\%). Enc.Total and Dec.Total denote total time for encoding and decoding.``\Checkmark'' and ``\XSolidBrush'' represent with and without this module, respectively.}
\label{Table:ab_ACGC}
\centering
\begin{tabular}{ccccccc}
\hline
Hyper-prior & AC & AG & Enc.Total(s) & Dec.Total(s) & BD-Rate(\%)  \\ 
\hline 
\Checkmark & \XSolidBrush & \XSolidBrush &0.22&0.29& 7.46	\\ 
\Checkmark & \Checkmark & \XSolidBrush 	&0.23& 0.30 & -4.11\\ 
\Checkmark & \XSolidBrush & \Checkmark & 0.26 & 0.32 & -0.19	\\ 
\Checkmark & \Checkmark & \Checkmark &0.27 & 0.34 & -6.27	\\ 
\hline
\end{tabular}
\end{table}

\textbf{Analysis of ACGC module.}
We conduct ablation experiments on the proposed ACGC module to study the impact of AC and AG components. We first remove the ACGC module, retaining only the hyperprior parameters. Then, we sequentially add other components. We used VTM-17.1 as the anchor (BD-Rate=0\%) to compare the encoding and decoding times as well as the BD-Rate among different schemes. The experimental results are presented in Table \ref{Table:ab_ACGC}. The results show that upon removing the ACGC module and relying on hyperprior parameters as the decoding context, the BD-Rate increased by 7.46\%. This shows that it achieves worse performance than VTM. Upon adding the AC, AG and ACGC modules, the BD-Rate decreased by 4.11\%, 0.19\% and 6.27\%, respectively. It is noteworthy that despite adding different components, there is no significant increase in encoding and decoding times. Thus, the ACGC context model demonstrated state-of-the-art performance.

\begin{figure}[!t]  
\centering  
\subfloat{  
\begin{minipage}[b]{0.3\linewidth}  
\centering  
\includegraphics[width=\columnwidth]{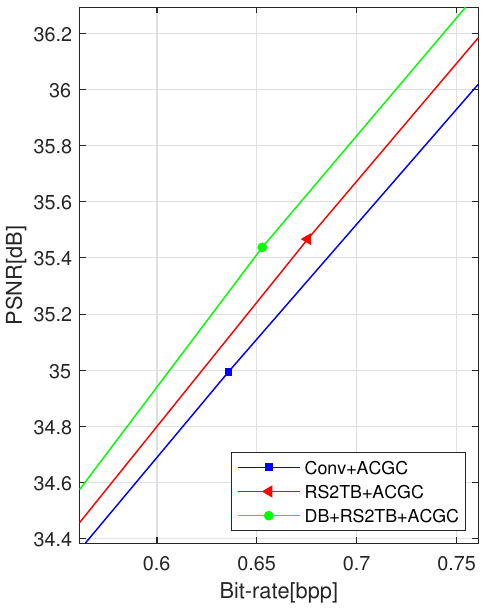}  
\end{minipage}  
}%
\subfloat{  
\begin{minipage}[b]{0.29\linewidth}  
\centering  
\includegraphics[width=\columnwidth]{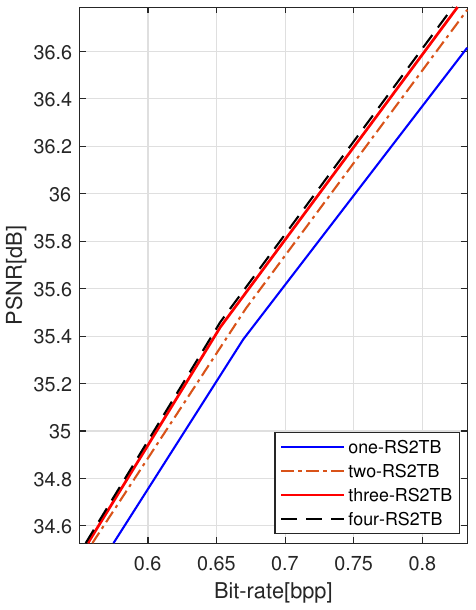}  
\end{minipage}  
}%
\caption{Ablation study of analysis transform module. The left is the performance of various components within ACGC module (``DB'' represents the enhancement module with the dense block, ``Conv'' and ``RS2TB'' are CNN-based and Residual SwinV2 Transformer-based models, respectively.), and the right is the performance of different quantities of RS2TB in the analysis module.}  
\label{Fig:ab_trans}
\end{figure}

\textbf{Analysis of analysis transform module.} 
In S2LIC, we replace traditional stacked residual blocks with SwinV2 transformer to achieve a non-linear transformation of images. We first show the different components of the the analysis transform module. Under the condition that the other parameters of the model are the same, we compare three models using ``CNN-based'', ``Residual SwinV2-based'' and ``DB + Residual SwinV2-based'', respectively. As shown in Fig. \ref{Fig:ab_trans} (Left), experimental results indicate that SwinV2 attention performs better in capturing global feature information compared to CNN-based models, resulting in a 0.11 dB improvement in PSNR. Additionally, the feature enhancement module based on the DB block leads to an increase by approximately 0.17-0.2 dB, thereby improving rate-distortion performance.

Furthermore, we conduct a detailed comparison of four different quantities of RS2TB in the main encoder, with the specific results shown in Fig. \ref{Fig:ab_trans} (Right). With only one RS2TB configured, the performance is poor. Increasing to two RS2TBs improves performance by approximately 0.2 dB. However, adding more RS2TBs does not lead to significant further improvement in performance when the number reaches four. Instead, it results in a substantial increase in model complexity and computation time. Therefore, to balance performance and complexity, we choose three RS2TBs as the primary transformation modules to achieve optimal compression results.

\section{Conclusion}
\label{Conclusion}
In this paper, we propose the ACGC model to efficiently achieve dual feature aggregation in both inter-slice and intra-slice contexts. The ACGC model is incorporated in a parallel checkerboard context model to achieve faster decoding speed and better rate-distortion performance. In addition, we also incorporate residual Swinv2 transformer block and a nonlinear feature enhancement module in the main encoder and main decoder networks to further reduce the spatial redundancy of the latent representations. The experimental results demonstrate our method achieves better performance than the best traditional codec VTM-17.1 and some recent learning-based image compression methods in both PSNR and MS-SSIM metrics. In future work, we will design more efficient and effective network frameworks to enhance rate-distortion performance. Additionally, we will reduce encoding and decoding times by designing more efficient and parallelizable entropy models.

\section{Acknowledgment}
\label{Acknowledgment}
This work was supported by the Aeronautical Science Foundation of China (Grant No. 20184370009), the Natural Science Foundation of Shaanxi Province, China (Grant No. 2021GXLH-Z081), the China Scholarship Council (No. 202306280309).

\bibliographystyle{elsarticle-num} 
\bibliography{ref}





\end{document}